\documentclass[useAMS,usenatbib]{mn2e}

\usepackage{graphicx}
\usepackage{times}
\usepackage{natbib}
\usepackage{amsmath}

\newcommand{\gtrsim}{\ga}
\newcommand{\lesssim}{\la}

\def\om{ \Omega_{ {\rm 0,M} } }
\def\ob{ \Omega_{ {\rm 0,b} } }
\def\ol{ \Omega_{ {\rm 0,\Lambda} } }
\def\msun{ {\rm M_\odot} }
\def\vbx{ v_{\rm b,x} }

\title[Gas bulk flows]{The impact of primordial supersonic flows on early structure formation, reionization and 
the lowest-mass dwarf galaxies}
\author[Maio, Koopmans \& Ciardi]{
Umberto~Maio$^{1}$\thanks{umaio@mpe.mpg.de},
Leon~V.~E.~Koopmans$^{2}$, and
Benedetta~Ciardi$^{3}$\\~\\
${}^1$ Max Planck Institute for Extraterrestrial Physics, Giessenbachstra{\ss}e 1, 85741 D-Garching, Germany\\
${}^2$ Kapteyn Astronomical Institute, University of Groningen, P.O.Box 800, 9700AV, Groningen,
the Netherlands\\
${}^3$ Max Planck Institute for Astrophysics, Karl-Schwarzschild-Stra{\ss}e 1, 85741 D-Garching,Germany}

\begin{document}

\date{\today}
\pagerange{\pageref{firstpage}--\pageref{lastpage}}\pubyear{0}
\maketitle
\label{firstpage}

\begin{abstract}
Tseliakhovich \& Hirata recently discovered that higher-order corrections to the cosmological linear-perturbation theory lead to supersonic coherent baryonic flows just after recombination (i.e.\ $z \approx 1020$), with rms velocities of $\sim$30 km/s relative to the underlying dark-matter distribution, on comoving scales  of $\la 3$~Mpc\,$h^{-1}$.
To study the impact  of these coherent flows we performed high-resolution N-body plus SPH simulations in boxes of 5.0 and 0.7 Mpc\,$h^{-1}$, for bulk-flow velocities of 0 (as reference), 30 and 60 km/s. 
The simulations follow the evolution of cosmic structures by taking into account detailed, primordial, non-equilibrium gas chemistry (i.e.\ H, He, H$_2$, HD, HeH, etc.), cooling, star formation, and feedback effects from stellar evolution.
We find that these bulk flows suppress star formation in low-mass haloes (i.e.\ $M_{\rm vir} \la 10^8$M$_{\odot}$ until $z\sim 13$), lower the abundance of the first objects by $\sim 1\%-20\%$, and as consequence delay cosmic star formation history by $\sim 2\times 10^7\,\rm yr$.
The gas fractions in individual objects can change up to a factor of two at very early times. 
Coherent bulk flow therefore has implications for (i) the star-formation  in the lowest-mass haloes (e.g. dSphs), (ii) the start of reionization by suppressing it in some patches of the Universe, and (iii) the heating (i.e.\ spin temperature) of neutral hydrogen. 
We speculate that the patchy nature of reionization and heating on several Mpc scales could lead to enhanced differences in the HI spin-temperature, giving rise to stronger variations in the HI brightness temperatures during the late dark ages.
\end{abstract}



\section{Introduction}\label{Sect:introduction}

Cosmic structure-formation models are based on Jeans theory \cite[][]{Jeans1902} applied to primordial matter fluctuations in an expanding Universe \cite[e.g.][]{wmap7_2010}.
The matter density perturbations are linearly expanded to first order, which is supposed to work well as long as the density contrast is much smaller than unity \cite[e.g.][]{Peebles1974,ColesLucchin2002,CiardiFerrara2005}, and the equations that govern the evolution of the dark and baryonic matter become linear.
To study highly non-linear evolution when the density contrast approaches and far exceeds unity, however, fully numerical simulations are required.
According to the concordance cosmological model \citep[e.g.][]{wmap7_2010} the Universe is flat and has an expansion rate of $H_0\simeq 71\,\rm km/s/Mpc$. The cosmological constant (or dark energy) represents the main energy-density content, $\ol\simeq 0.7$, whereas dark matter contributes $\om\simeq 0.26$ and baryonic matter is only a small fraction of $\ob\simeq 0.04$.
Within this context, baryonic structures arise from in-fall and condensation of gas into growing dark-matter potential wells \cite[e.g.][]{GunnGott1972,Peebles1974,WhiteRees1978}.
Both in-fall and condensation of gas only take place {\sl after} the time of recombination ($z\simeq 1020$), when baryons and photons decouple \cite[][]{wmap7_2010}.
On the other hand, dark matter perturbations can already start growing when their density exceeds that of radiation (i.e.\ at $z_{\rm eq}\sim 3100-3200$).
Recently \citet[][]{TseliakhovichHirata2010} for the first time included quadratic terms in the cosmological perturbation theory to account for the advection of small-scale perturbations by large-scale velocity flows.
They find that, at the redshift of decoupling $z\simeq 1020$, coherent supersonic flows of the baryons relative to the underlying dark-matter distribution are formed on scales of a few Mpc or less, with typical velocities on those 
scales with a rms value of $\sim 30\,\rm km/s$, sourced by density perturbations up to scales of $\sim$100\,Mpc.
These gas bulk velocities could actually suppress the formation of the very early and low-mass structures, induce higher baryon acoustic oscillations \cite[][]{Dalal_et_al_2010},  and have further implications on high-redshift galaxy clustering, 21-cm studies, and reionization. In fact, as we find, they might impact the formation of the lowest-mass galaxies (e.g.\ dSph) on scales as large as the local group, depending of the magnitude of the bulk flow at decoupling.
The effect can thus cause a spatially varying bias in baryonic structure formation, whereas small-scale baryonic structures are suppressed more in regions where the bulk-velocity is larger.
Detailed numerical simulations considering this non-linear effect are not yet available and the very high dynamic range required is not foreseen in the near future.
It therefore looks very challenging to quantitatively assess these effects in the highly non-linear regime discussed by \cite{TseliakhovichHirata2010}.
Indeed, one would need to resolve, at the same time, large scales (of the order of the horizon) and extremely small scales (of the order of the Jeans length).
Given these extreme difficulties, we adopt the following approach: we focus on small scales where the effect is expected to be largest,  at least properly dealing with all the detailed physics and chemistry, and we assume {\sl a constant bulk velocity shift in the initial conditions} as determined by \cite{TseliakhovichHirata2010} on the relevant scales. 
Even though this approximation is still crude, we believe it is a reasonable assumption to study the lowest-mass structures in the box (which have virial radii much smaller than the box size), we expect it to be able to provide the first quantitative results in the highly non-linear regime and assess its effects on early structure formation. 
In this paper, we make a first step towards understanding the effects of coherent supersonic bulk flow on the formation of the smallest baryonic structures, in particular focusing on the suppression of star formation and the delay of reionization. In Sect. \ref{sect:sims}, we describe our simulations, and in Sect. \ref{sect:results} we show our results. We summarize and conclude in Sect. \ref{sect:conclusion}.

\section{Simulations}\label{sect:sims}

We perform numerical simulations by using the parallel tree/SPH P-Gadget2 code \cite[][]{Springel2005}, which implements gravity, hydrodynamics, molecular and atomic evolution, cooling, population III and population II star formation, supernova/wind feedback, and a full chemistry network involving e$^-$, H, H$^+$, H$^-$, He, He$^+$, He$^{++}$, H$_2$, H$_2^+$, D, D$^+$, HD, and HeH$^+$ \cite[for more details see][]{Maio2006,Maio2007,Maio2010,MaioPhD,Maio_et_al_2010b}.
Particularly important for our purposes is the capability of the code of following gas collapse down to the catastrophic molecular cooling branch, by resolving the Jeans scales of primordial structures with $\sim 10^2$ particles \cite[][]{Maio2009}, and completely fulfilling the correct resolution requirements for SPH simulations \cite[see][ for details]{BateBurkert1997,Maio2009}.
We generate the initial conditions by using the N-GenIC code and add bulk velocity shifts for the gas 
in the $x$-direction $\vbx = 0, 30, 60 \,\rm km/s$ (referred to as HR-Shift00, HR-Shift30, 
and HR-Shift60, respectively) at recombination epoch and properly scaled down 
\cite[according to][]{TseliakhovichHirata2010} to the simulation initial redshift of $z_{in}=100$.
The value $\vbx\sim 60\,\rm km/s$ has been considered to account for statistical variations from the expected
rms value of $\sim 30\,\rm km/s$, but it should be considered as an upper limit which can be found in less than 1\%
of the volume of the Universe.
Even though we expect that the presence of bulk flows at higher redshift does not affect the structure 
formation process, we nevertheless run additional simulations with $z_{in}=1020$, 
with $\vbx=0,60\,\rm km/s$ (HR-Shift00Rec, HR-Shift60Rec).
We should caution the reader that the Gadget code implements N-body/SPH structure formation, 
cooling, star formation, and feedback effects in matter-dominated universes, while
contributions from radiation or from the primordial hot plasma in the cosmological dynamics 
are neglected. Therefore, its use at $z >> 200$ might be questionable.
In any case, the results are very similar to the simulations with $z_{in}=100$, and thus
we don't discuss them any further.
Our reference simulations have a box side of $0.7\,\rm Mpc/{\it h}\simeq 1~Mpc$, and are run 
in the frame of the standard $\Lambda$CDM cosmological model ($\ol=0.7$, $\om=0.26$, $\ob=0.04$, $h=0.7$, $\sigma_8=0.9$, $n=1$).
Matter and gas fields are sampled with 320$^3$ particles, respectively, allowing a resolution of $\sim 10^2\msun/h$ for the gas component and $\sim 7.5\times 10^2 \msun/h$ for the dark-matter component.
The chemical abundances are initialized as in \citet{Maio2007,Maio2010}.
To check if the effects of primordial supersonic gas flows on structure formation could be 
seen in bigger but lower-resolution simulations, we also run a set of three larger boxes 
(LR-Shift00, LR-Shift30, and LR-Shift60) with a size of 5~Mpc/{\it h}, $\vbx=0, 30, 60,\rm km/s$, 
and with the same parameters described before. This box is comparable to the smallest scale
where the bulk flows are expected according to \cite{TseliakhovichHirata2010}.
The corresponding mass resolution is $\sim 4\times 10^4\msun/{\it h}$ and $\sim 3\times 10^5\msun/{\it h}$, for gas and dark-matter species.
In this case, we did not find any relevant differences among the three runs, implying that the effect is observable {\sl only} if the numerical resolution is high enough to resolve halo masses of $\sim 10^4\msun/{\it h}$.
This highlights the difficulty of current large-scale simulations to address these effects.
\\
In Table \ref{tab:sims} we list the main features of the numerical set-ups.
\begin{table*}
\centering
\caption[Simulation parameters]{Parameters adopted for the simulations.}
\begin{tabular}{lccccccc}
\hline
\hline
Model & box side & number of & mean inter-particle & $M_{gas}\rm [M_\odot/{\it h}]$ & $M_{dm}\rm [M_\odot/{\it h}]$ & initial & bulk shift [km/s]\\
& [Mpc/{\it h}] & particles & separation [kpc/$h$] & & & redshift & at $z=1020$ \\
\hline
HR-Shift00 & 0.7 &$2\times 320^3$ & 2.187 & $1.16\times 10^2$ & $7.55\times 10^2 $ & 100 & 0\\
HR-Shift30 & 0.7 &$2\times 320^3$ & 2.187 & $1.16\times 10^2$ & $7.55\times 10^2 $ & 100 & 30\\
HR-Shift60 & 0.7 &$2\times 320^3$ & 2.187 & $1.16\times 10^2$ & $7.55\times 10^2 $ & 100 & 60\\
HR-Shift00Rec & 0.7 &$2\times 320^3$ & 2.187 & $1.16\times 10^2$ & $7.55\times 10^2 $ & 1020 & 0\\
HR-Shift60Rec & 0.7 &$2\times 320^3$ & 2.187 & $1.16\times 10^2$ & $7.55\times 10^2 $ & 1020 & 60\\
\hline
LR-Shift00 & 5.0 &$2\times 320^3$ & 15.62 & $4.23\times 10^4$ & $ 2.75\times 10^5$ & 100 & 0\\
LR-Shift30 & 5.0 &$2\times 320^3$ & 15.62 & $4.23\times 10^4$ & $ 2.75\times 10^5$ & 100 & 30\\
LR-Shift60 & 5.0 &$2\times 320^3$ & 15.62 & $4.23\times 10^4$ & $ 2.75\times 10^5$ & 100 & 60\\
\hline
\label{tab:sims}
\end{tabular}
\begin{flushleft}
\vspace{-0.5cm}
\end{flushleft}
\end{table*}

\section{Results}\label{sect:results}

We start  by looking at the overall evolution of the star formation rate density of the Universe, and then 
we consider the main statistical quantities involved in the cosmological evolution process, i.e. the 
distributions of dark-matter haloes and gas clouds.

\subsection{Star formation rate density}

\begin{figure}
\centering
\includegraphics[width=0.40\textwidth]{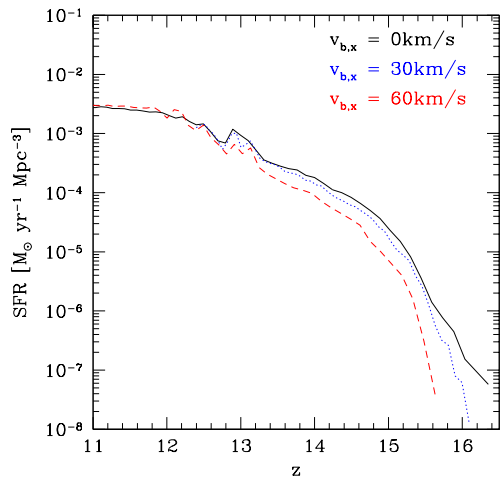}
\caption[SFR]{\small
Total star formation rates for bulk velocity shifts of
$0\,\rm km/s$ (black solid line),
$30\,\rm km/s$ along the x-axis (blue dotted line),
$60\,\rm km/s$ along the x-axis (red dashed line)
are shown for the simulation box of side $\sim 1$~Mpc.
}
\label{fig:sfr}
\end{figure}

The behavior of the star formation rate densities is plotted in Fig. \ref{fig:sfr}, for all the cases considered before.
What emerges is a systematic shift in the onset of star formation, with a delay increasing for higher values of $\vbx$.
For the $\vbx=0\,\rm km/s$ case, the onset happens at $z\sim 16.3$, when the Universe is roughly $2.5\times 10^8\,\rm yr$ old, while in the $\vbx=30\,\rm km/s$ and $60\,\rm km/s$ cases, star formation sets in at $z\sim 16.1$ and 15.6, respectively, corresponding to a delay of $\sim 10^7\,\rm yr$ and $\sim 2\times 10^7\,\rm yr$ compared to the case without velocity shift.
Differences at very early times are almost one order of magnitude and at later times
 a factor of a few, persisting down to $z\sim13$, when the global trends rapidly converge to similar values.
The reason why this happens is due to the higher kinetic energy given to the gas at the recombination epoch: because of that, first dark-matter haloes can retain less material, gas condenses more slowly hindering molecule formation, the corresponding cooling is less efficient, and the resulting star formation is delayed.
In particular, this is very relevant for small primordial haloes, with masses of $\sim 10^4\msun - 10^8\msun$ \cite[see also][]{TseliakhovichHirata2010}, whose dimensions are comparable to or smaller than the baryon Jeans length.
In this case in fact, the gas cannot fragment within the halo and partially or entirely flows out of it, since larger dark-matter potential wells are needed to retain the gas.
Therefore, this effect on small scales influences the growth of larger objects, as well.
From this we conclude that the inflow and condensation of gas, and thus also star formation, is already strongly suppressed in haloes in the mass-range $\la 10^8\msun$ {\sl before reionization}. Those low-mass halos surviving to the present day, or even those that merge into larger masses, would therefore have a lower baryonic and stellar mass fraction.
We further note that the effect of suppressing star formation is similar to that of reionization, but it has a very different physical origin and occurs well before reionization itself.

\subsection{Dark-matter haloes and gas clouds}
\begin{figure*}
\centering
\includegraphics[width=0.22\textwidth]{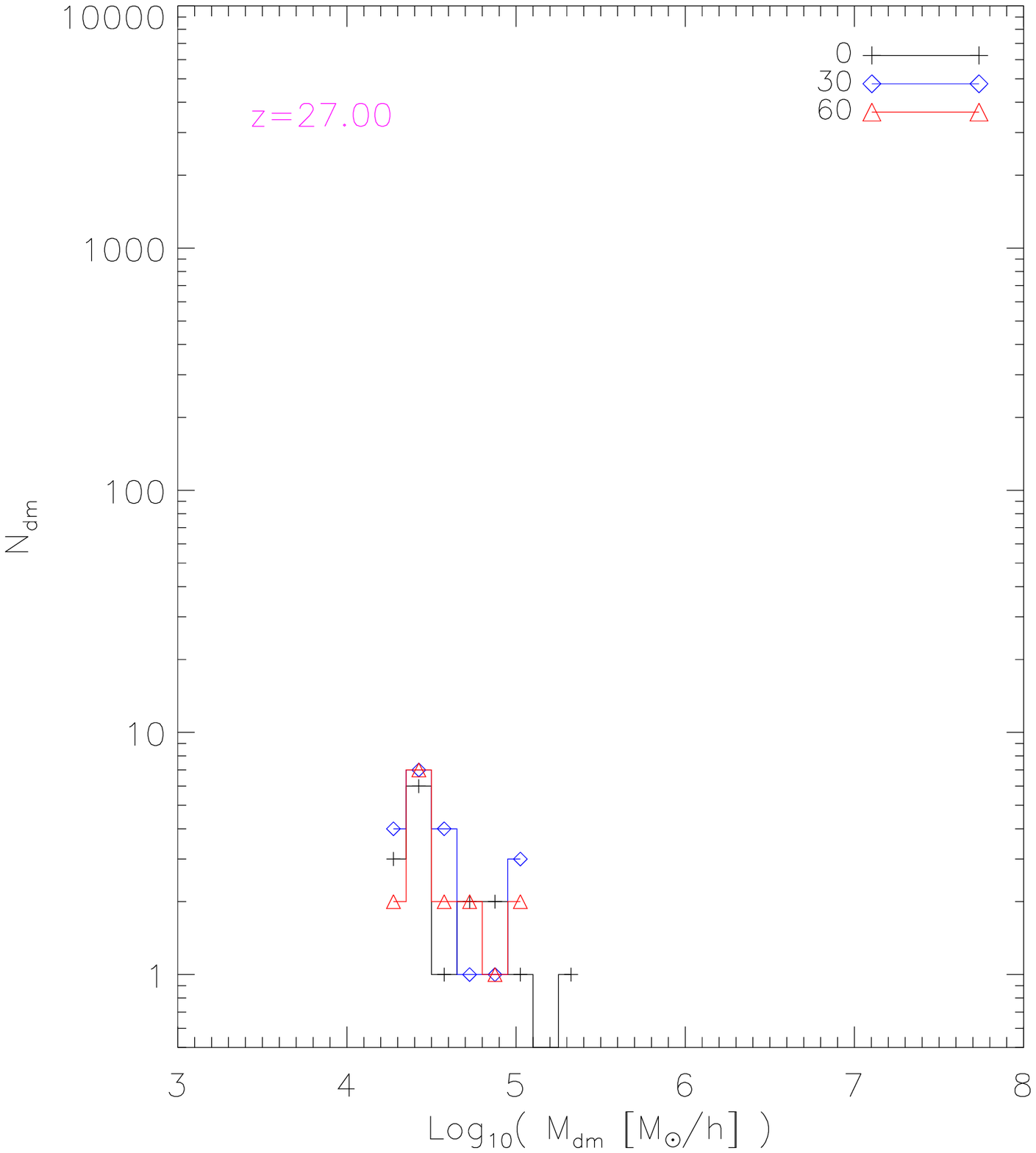}
\includegraphics[width=0.22\textwidth]{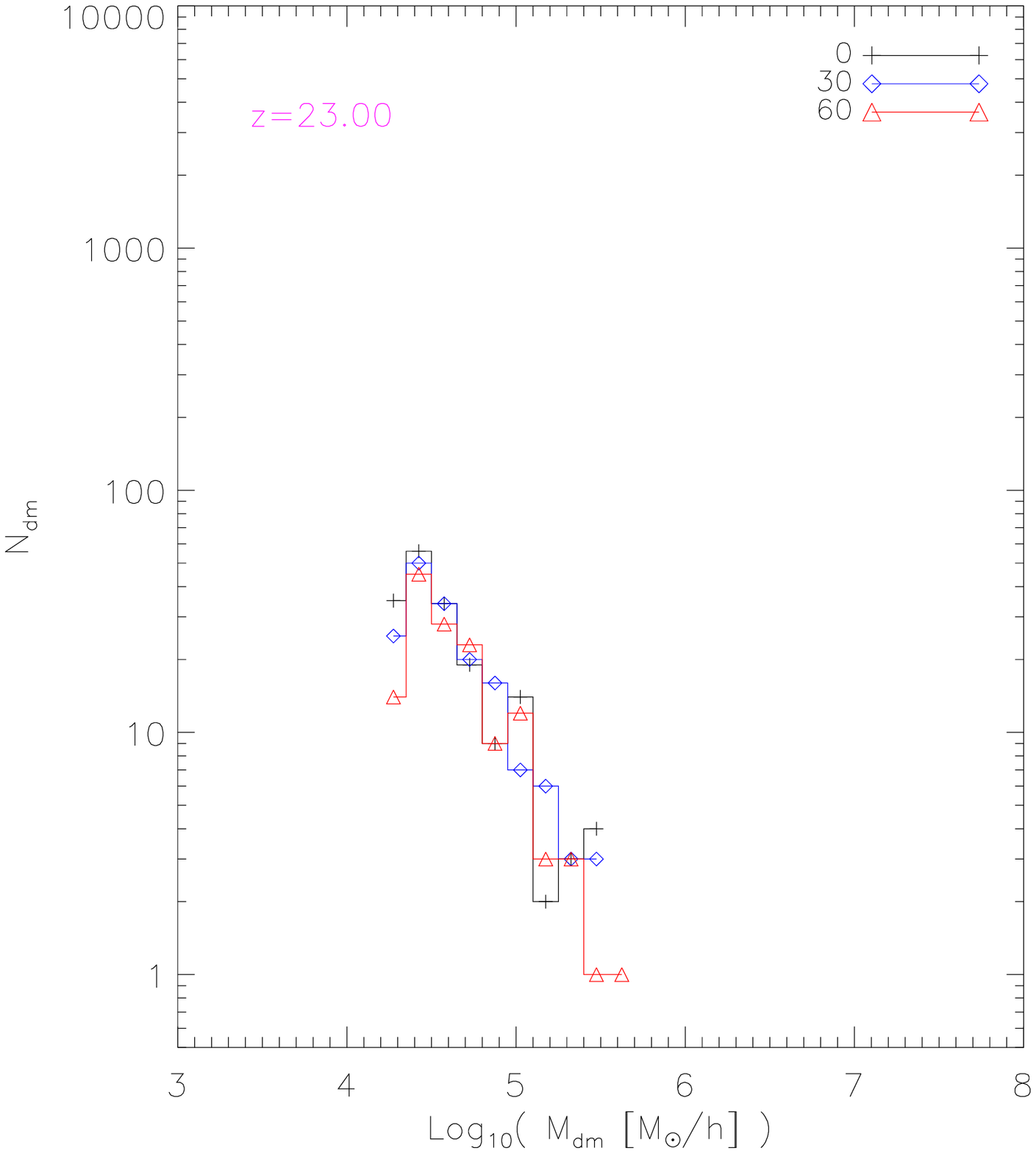}
\includegraphics[width=0.22\textwidth]{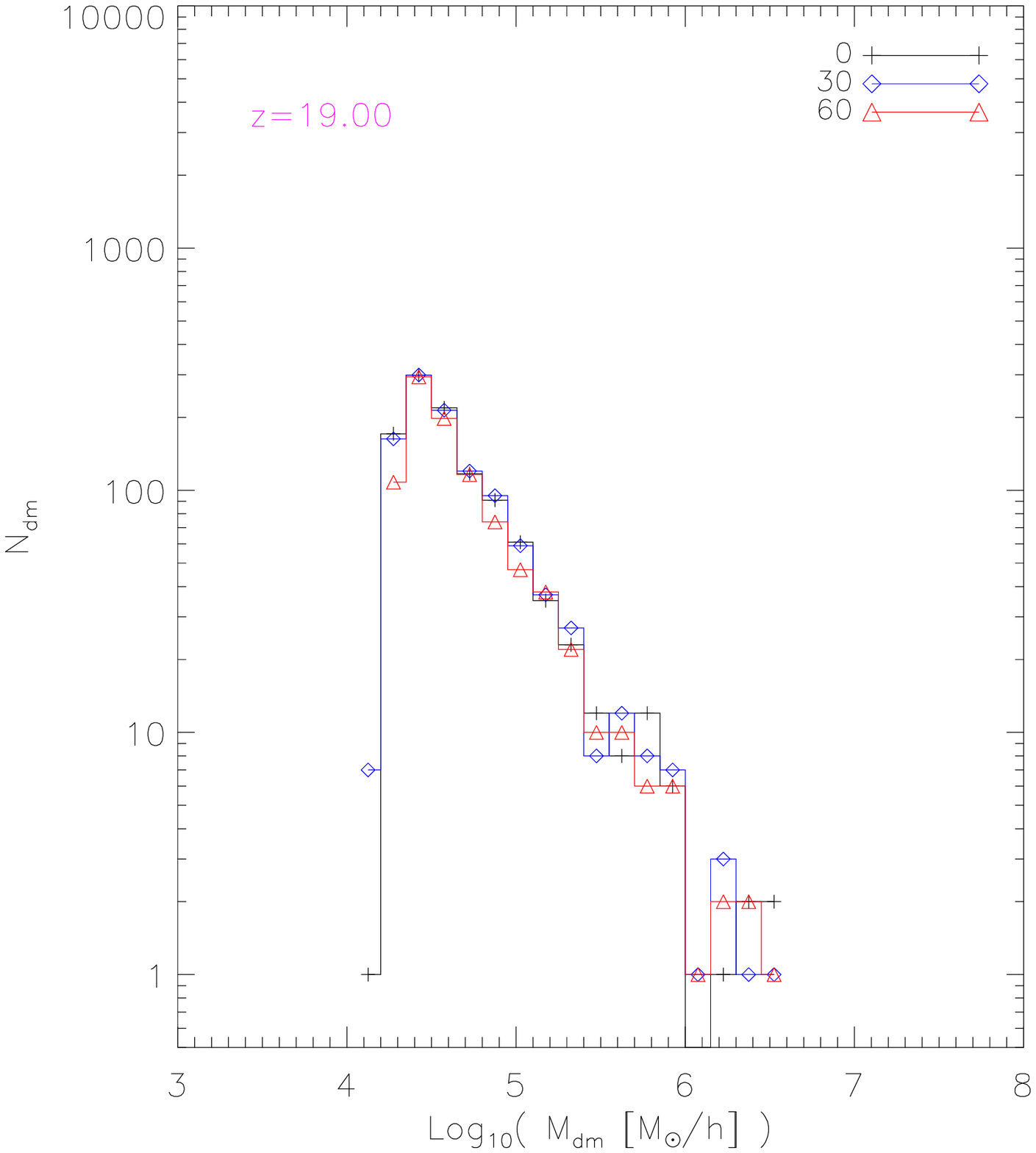}
\includegraphics[width=0.22\textwidth]{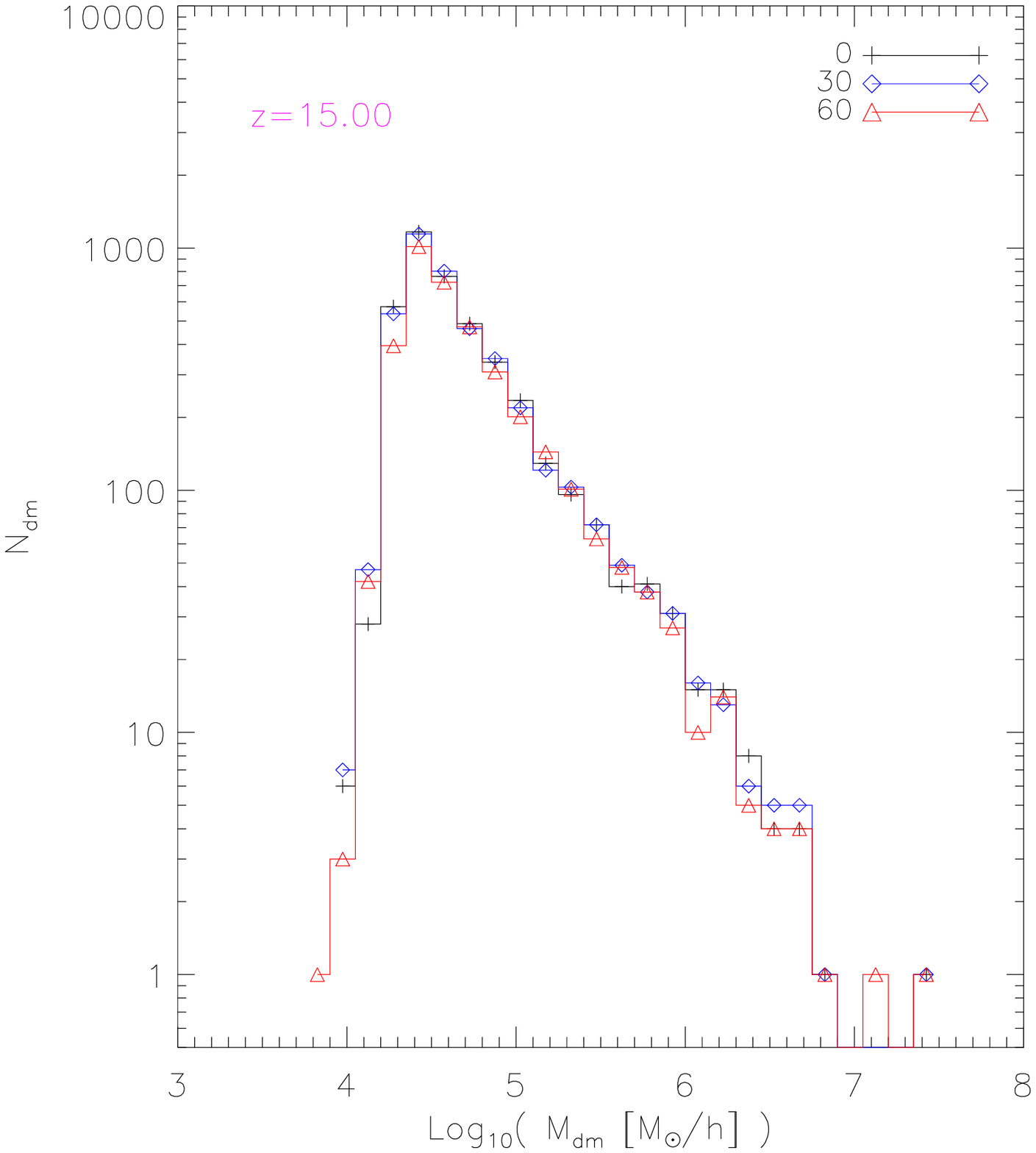}
\centering
\includegraphics[width=0.22\textwidth]{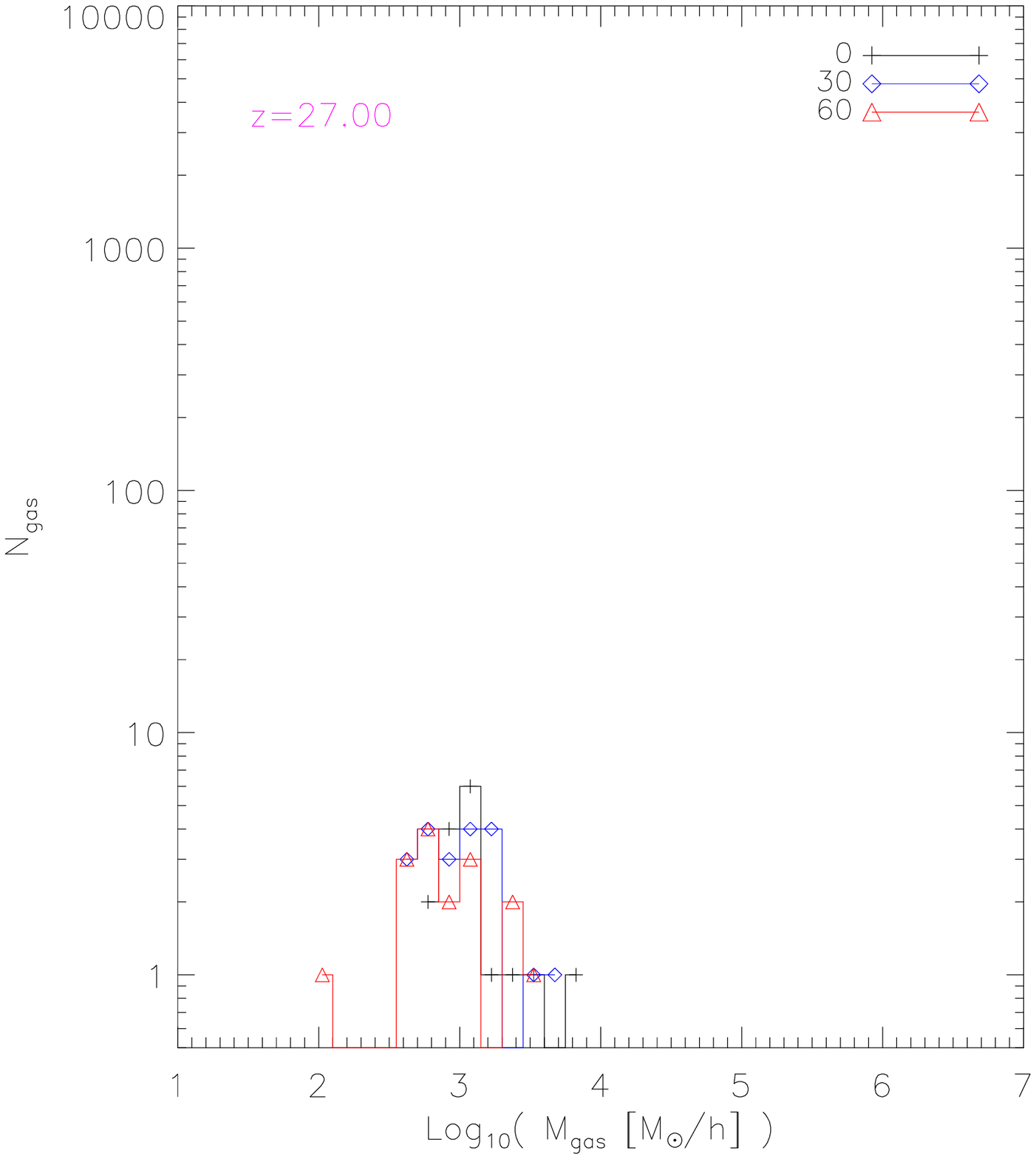}
\includegraphics[width=0.22\textwidth]{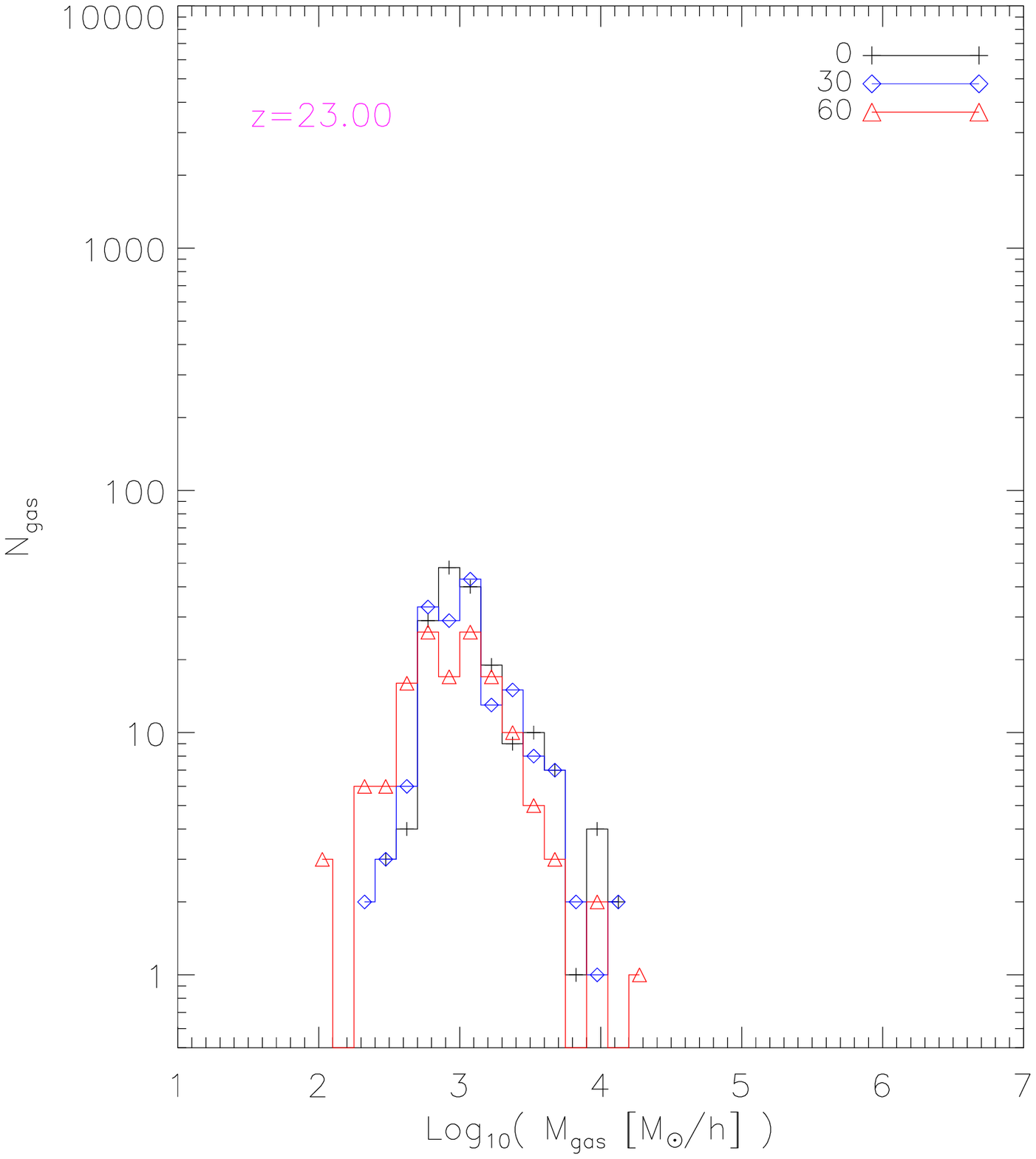}
\includegraphics[width=0.22\textwidth]{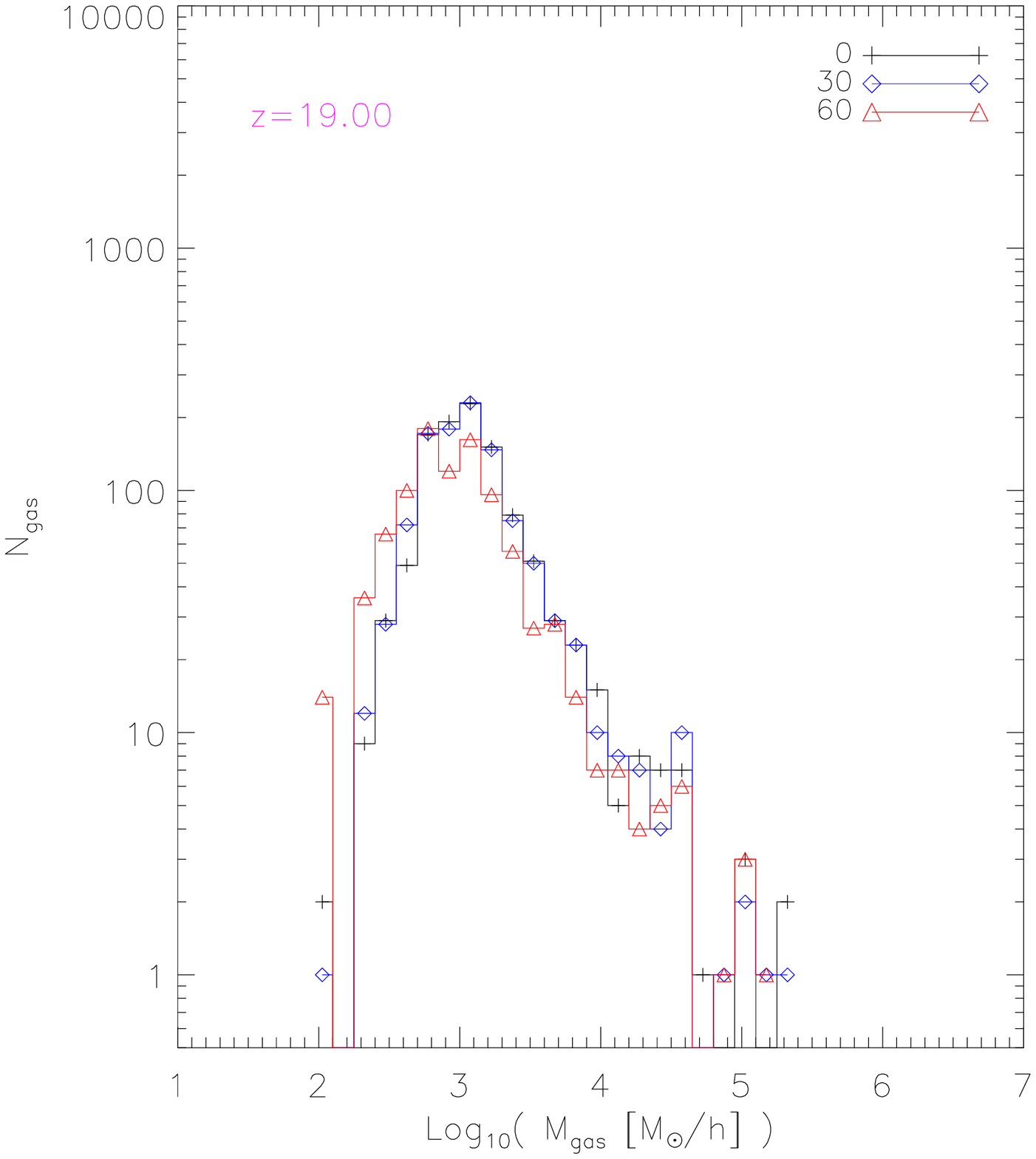}
\includegraphics[width=0.22\textwidth]{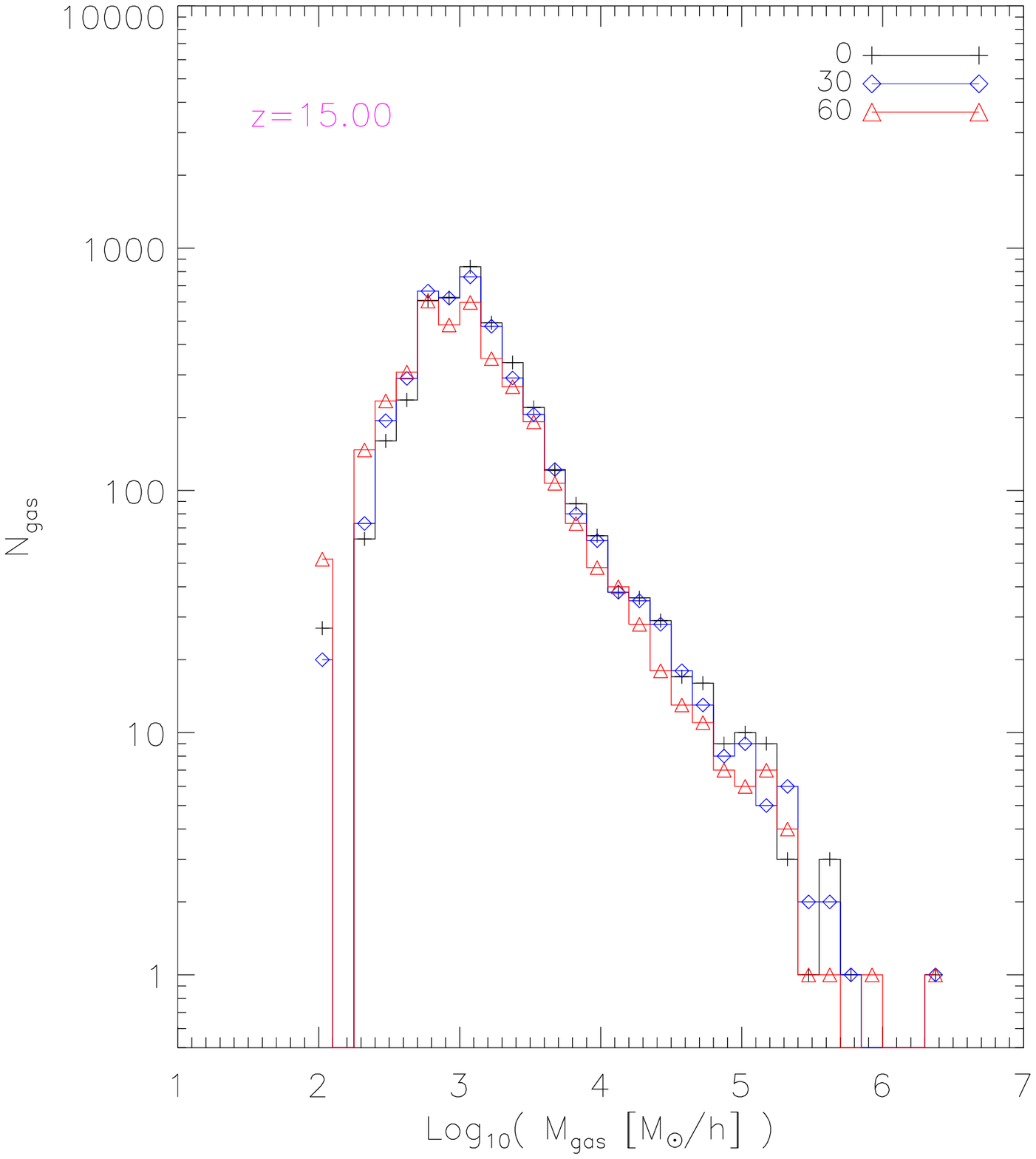}
\caption{{\it Top row}: Dark-matter halo distributions at different redshifts (see labels) for the 1~Mpc side boxes.
{\it Bottom row}: Gas cloud distributions at different redshifts (see labels) for the 1~Mpc side boxes.
Histograms refer to velocity shifts of
$0\,\rm km/s$ (black crosses),
$30\,\rm km/s$ along the x-axis (blue rhombi),
$60\,\rm km/s$ along the x-axis (red triangles).
}
\label{fig:haloes07}
\end{figure*}

In each simulation, we identify cosmic structures with the use of a friend-of-friend (FoF) algorithm. Each object is represented by all those particles (dark-matter, gas and star) closer than a limiting linking length of $20\%$ the mean inter-particle separation.
The dark-matter halo and gas cloud distributions are shown in Fig. \ref{fig:haloes07}, at different redshift as indicated by the legends, for the runs starting at $z_{in}=100$.
In general, at very high redshift we cannot give solid conclusions, since small-number statistic dominates the sample.
At later times (i.e.\ $z\lesssim 23$), some systematic behavior clearly emerges.
The mass distributions of dark matter haloes range between $\sim 10^4\msun/h$ and $10^7\msun/h$.
The three cases of $\vbx = 0, 30, 60\,\rm km/s$ show a slight decrement with increasing $\vbx$, but only at a few percent level.
Hence, as expected, the effect of the bulk-flow of gas on the overall growth of the dark-matter dominated haloes is minimal.
More clear differences arise from the analyses of the gas clouds.
In Fig.\ref{fig:haloes07}, we also show the mass distribution of the gaseous component of the objects and we clearly see decrements up to tens of percent, mostly evident in the $\vbx = 60\,\rm km/s$ case.
Interestingly, there is a cascade effect from smaller to larger masses (expected from the simple analyses of the star formation rates), which delays and slightly hinders the accumulation of gas in the dark-matter haloes (i.e.\ suppression of gas inflow in lower-mass haloes also suppresses the gas accumulation in high-mass haloes through accretion of low-mass haloes).
As expected, the average gas cloud mass decreases for larger $\vbx$.
At $z\simeq 23$, it drops from 
$\sim 1.70\times 10^3\msun/{\it h}$ ($\vbx=0\,\rm km/s$) down to 
$\sim 1.62\times 10^3\msun/{\it h}$ ($\vbx=30\,\rm km/s$), and 
$\sim 1.39\times 10^3\msun/{\it h}$ ($\vbx=60\,\rm km/s$).
While at $z\simeq 15$, it is 
$\sim 4.41\times 10^3\msun/{\it h}$ ($\vbx=0\,\rm km/s$), 
$\sim 4.26\times 10^3\msun/{\it h}$ ($\vbx=30\,\rm km/s$), and
$\sim 4.06\times 10^3\msun/{\it h}$ ($\vbx=60\,\rm km/s$).
The differences correspond to decrements with respect to the reference run of 
$\sim 6\%-3\%$ ($\vbx=30\,\rm km/s$) and $\sim 20\%-10\%$ ($\vbx=60\,\rm km/s$) at $z\simeq 23-15$, respectively.
Also the total abundance of gas clouds shows a similar trend, with mass-weighted number counts 
dropping, with respect to the reference run, by $\sim 5\%-3\%$ ($\vbx=30\,\rm km/s$) and 
$\sim 20\%-50\%$ ($\vbx=60\,\rm km/s$) at $z\simeq 23-15$.
We have verified that the excess small clumps observed in the low-mass tail of the 
distributions (at masses $\lesssim 500\msun/{\it h}$) for $\vbx>0$ are due to residual gas which,
differently from the reference run, is not trapped by the dark-matter haloes
and, later collapses to form such clumps. It should be noted though that
statistical errors in this mass regime are substantial, due to the small number of particles contained in a halo.
\begin{figure}
\centering
\includegraphics[width=0.40\textwidth]{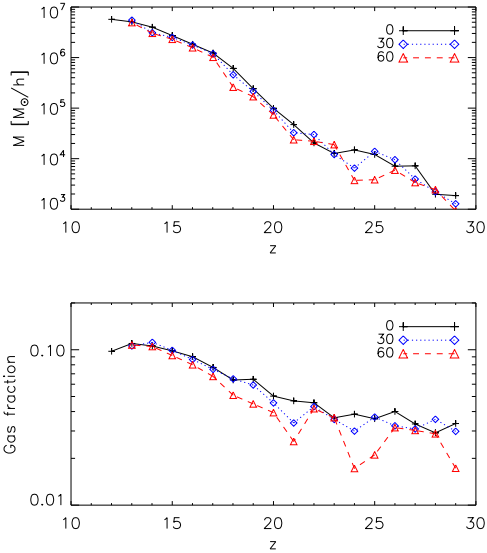}
\caption[Gas Evolution (z=100)]{\small
Evolution of the gas content (top) and of the gas fraction (bottom) for the first (and most massive) halo formed in the simulations.
Data refer to velocity shifts of 
$0\,\rm km/s$ (solid lines, black crosses),
$30\,\rm km/s$ (dotted lines, blue rhombi),
$60\,\rm km/s$ (dashed lines, red triangles).
}
\label{fig:evolution07}
\end{figure}
As a more specific example, in Fig. \ref{fig:evolution07} we show the evolution of the gas content (top panel) and gas fraction (bottom panel) of the first (and most massive) halo formed in the simulations.
The gas mass ranges between $\sim 10^3\msun/h$ and $10^7\msun/h$, and the corresponding dark-matter mass is about one order of magnitude larger.
The gas mass variations due to the different values of $\vbx$ are significant at $z\gtrsim 16$, while at smaller redshift the total amount converges to $\sim 2-6\times 10^6\msun/h$.
In this respect, the gas fractions in the halo are more informative, and suggest a depletion of gas for higher $\vbx$ up to a factor of 2.
Indeed, at $z\gtrsim 24$, the gas fraction for the $\vbx=0\,\rm km/s$ case is $\sim 0.04$, while in the $\vbx=30\,\rm km/s$ and $\vbx=60\,\rm km/s$ cases it drops to $\sim 0.03$ and below $0.02$, respectively.
Similar strong variations are observed also at later times (e.g. at $z\sim 23$), but the general trend converges at $z\lesssim 16$, when the discrepancies among the cases become $< 10\%$ and all the gas fractions reach values of $\sim 0.10$.
The same effects are observed in the gas mass and density profiles as function of the (physical) radius -- see Fig. \ref{fig:profiles07} and compare to Fig. 2 in \cite{TseliakhovichHirata2010}.
At early times (left panels) the amount of gas falling in the halo depends on $\vbx$, and the  larger $\vbx$, the more gas is depleted in the distribution tails. At larger radii, the mass depletion is up to $\sim 50\%$ and smoothly 
decreases towards the core, where the gas is even denser and more abundant.
This can be seen as a consequence of the fact that there is less gas falling in, and it sinks more easily in the potential wells of the halo.
Of course, given the low number of SPH particles reaching those regimes, the statistical errors might 
be quite significant and ad hoc simulations are needed to address this issue in more detail. 


\section{Summary and conclusions}\label{sect:conclusion}
\begin{figure*}
\centering
\includegraphics[width=0.45\textwidth]{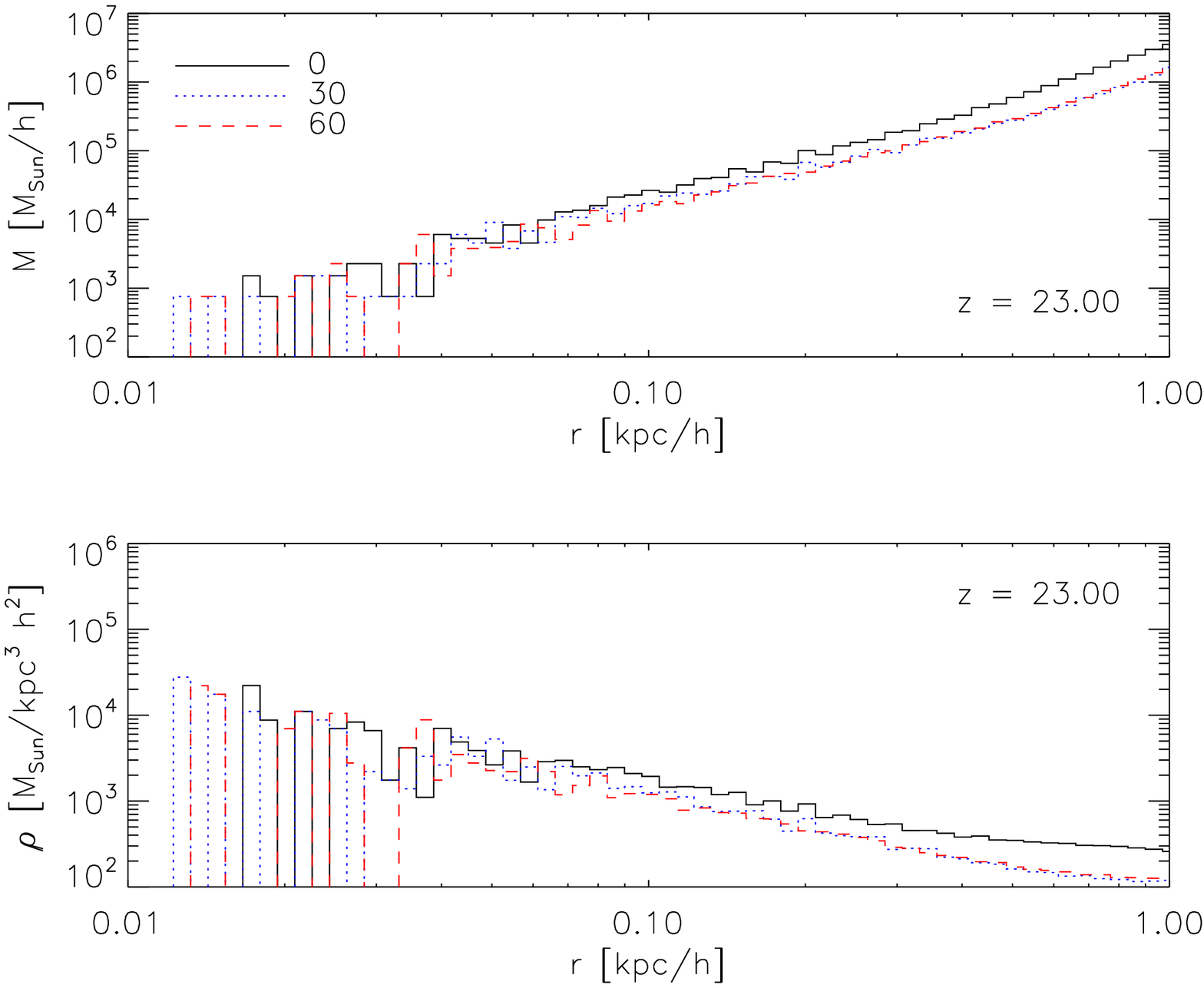}
\includegraphics[width=0.45\textwidth]{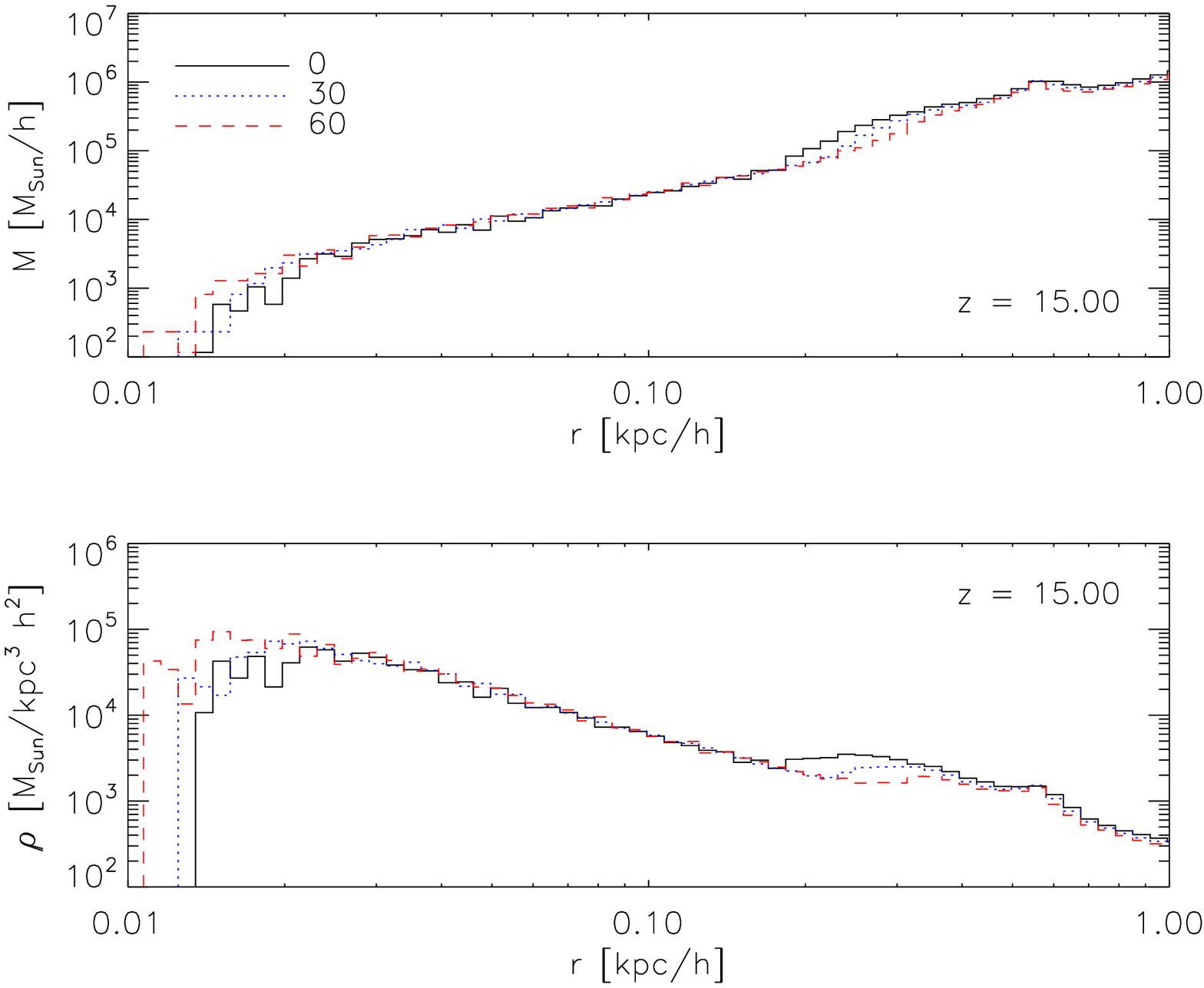}
\caption[Profiles (z=100)]{\small
Gas profiles as a function of the (physical) radius at redshift $z=23$ (left panels) and $z=15$ (right panels) 
for the data referring to the simulations with velocity shifts of 
$0\,\rm km/s$ (solid black lines),
$30\,\rm km/s$ (dotted blue lines) and
$60\,\rm km/s$ (dashed red lines).
Top panels show the gas mass profiles, while bottom panels show the gas density profiles.
}
\label{fig:profiles07}
\end{figure*}
Stimulated by very recent analytical work \cite[][]{TseliakhovichHirata2010}, we have run cosmological, high-resolution numerical N-body/SPH simulations including detailed, non-equilibrium chemistry evolution, cooling, star formation and feedback effects \cite[see e.g.][]{Maio2007,Maio_et_al_2010b}, which consistently follow the gas behavior down to the catastrophic cooling regime, and correctly resolve the small, primordial Jeans scales with $\sim 10^2$ SPH particles.
To take into account very early, supersonic, coherent, Mpc-scale flows, generated at recombination time from the advection of small-scale perturbations by large-scale velocity flows, we set initial gas velocity shifts along the $x$-direction, $\vbx=0,30,60\,\rm km/s$ based on analytic estimates of the rms differences in dark-matter and baryonic gas velocities at $z=1020$.
We find that in the simulations with larger $\vbx$ the onset of star formation is delayed by some $\sim 10^7\,\rm yr$, and it is smaller by a factor of a few in comparison to the $\vbx=0\,\rm km/s$ case, until redshift $z\sim 13$.
As a consequence, the beginning of the reionization process is delayed by the same amount and initially driven by more massive objects.
More ad hoc calculations should be done to quantify the differences in terms of evolution, topology and photon budget.
This might also have consequences in terms of detection of the 21cm signal from neutral hydrogen, because the building up of a Ly$\alpha$ background \cite[see e.g.][for a recent review]{2010Natur.468...49R}  necessary for the observability of the line \citep[e.g.][]{2006PhR...433..181F} would be delayed as well. We speculate that this could lead to more patchy reionization and HI heating, with larger HI brightness-temperature fluctuations. More research is needed to quantify these effects.
Because we find that the amount of gas in the first haloes is expected to drop significantly (up to a factor of 2 at early times, i.e. $\sim 50\%$), there could also be interesting consequences in terms of feedback effects, as e.g. less gas in small mass objects should make it easier to further remove gas via photoevaporation, winds or SN explosions. It will be interesting to investigate in more details the interplay between bulk-flow induced feedback effects.
Slight differences (at $\sim 1\%$ level) in the dark-matter statistical distributions also arise, but stronger differences (up above $\sim 10\%$ level) are seen in the gas clouds.
A cascade effect transfers the inability of primordial, small haloes to completely retain the gas to larger, lower-redshift haloes, whose gaseous components are systematically smaller for $\vbx > 0\,\rm km/s$.
In this way, the original hindering for the very first objects is transferred to the structure which form later on, affecting their growth history.
Numerical studies performed on lower-resolution simulations show no detectable differences among the different cases, and this means that the lack of numerical resolution can be a strong limitation to probe higher-order corrections to linear perturbation theory, and SPH particle masses of $\sim 10^2\msun/{\it h}$ are required.
The precise initial redshift does not alter significantly our findings either.
\\
Our general conclusions from this albeit preliminary, but {\sl first} fully non-linear, study of the effects of bulk gas flows on non-linear cosmic structure growth are that it:
(i) delays early star formation, which affects both reionization (by delaying it to lower redshifts) and the heating of gas of higher redshift, possible affecting total/global emission and absorption features of HI against the CMB by making these effects more patchy;
(ii) suppresses star formation in the lowest-mass haloes over all redshifts which has a cascading effect over nearly 300 million years  on higher-mass haloes which are their merger-products.
Because these small early objects, if they survive till the present day, are expected to be the parent population of e.g.\ dSph galaxies, the effct of the bulk flows can also have an impact on the stellar/gas content of the lowest-mass dwarf satellites around more massive galaxies (e.g.\ the MW) and be highly spatially dependent (i.e.\ the effect is only present where there were large bulk flows at decoupling).
This could result in a strongly spatially varying bias as suggested by \citet[][]{TseliakhovichHirata2010} and \citet[][]{Dalal_et_al_2010}.
In addition, suppression of gas infall and condensation on these small scales could play an important role in the ``missing satellite problem'' \cite[see e.g.][and references therein]{Kravtsov2010}.


\section*{acknowledgments}
We acknowledge the referee C.~Hirata, for his swift and postitive feeback.
We acknowledge useful discussions with J. Bolton, S. D. M. White and U. Pen.
UM acknowledges S. Khochfar and the tmox group.
LVEK acknowledges the generous support by an ERC-ST Grant.


\bibliographystyle{mn2e}
\bibliography{bibl.bib}

\label{lastpage}
\end{document}